\documentclass[showpacs,twocolumn,showkeys,amsmath,amssymb,pra]{revtex4-1}

\usepackage{color}
\usepackage{graphicx}   
\newcommand{\rd}{{\mathrm d}}
\newcommand{\re}{{\mathrm e}}
\newcommand{\ri}{{\mathrm i}} 
\newcommand{\ab}{a^{\phantom{\dagger}}}
\newcommand{\ad}{a^{\dagger}}                   

\begin{document}

\title[Quasiparticle tunneling]
      {Quasiparticle tunneling in a periodically driven bosonic Josephson 
       junction}

\author{Bettina Gertjerenken}
\altaffiliation[Present address: ]
       {Department of Mathematics and Statistics,
	University of Massachusetts,
	Amherst, MA 01003-4515, USA}
\author{Martin Holthaus}

\affiliation{Institut f\"ur Physik, Carl von Ossietzky Universit\"at, 
	D-26111 Oldenburg, Germany}
                  
\date{November 1, 2014}

\begin{abstract}
A resonantly driven bosonic Josephson junction supports stable collective 
excitations, or quasiparticles, which constitute analogs of the Trojan wave 
packets previously explored with Rydberg atoms in strong microwave fields. 
We predict a quantum beating effect between such symmetry-related many-body 
Trojan states taking place on time scales which are long in comparison with 
the driving period. Within a mean-field approximation, this quantum beating 
can be regarded as a manifestation of dynamical tunneling. On the full 
$N$-particle level, the beating phenomenon leads to an experimentally feasible,
robust strategy for probing highly entangled mesoscopic states.
\end{abstract} 




\maketitle 


\section{Introduction}
\label{sec:1}

The mechanism which effectuates the stability of ion motion in a Paul
trap~\cite{Paul90} also underlies the stability of the motion of the Trojan 
asteroids, which orbit around the Sun near stable Lagrange points of the
Sun-Jupiter system. This stable celestial motion has a quantum mechanical
counterpart, discovered in 1994 by Bialynicki-Birula, Kali\'{n}ski, and Eberly:
If one exposes Rydberg electrons to strong microwave radiation, such that the 
classical Kepler frequency of the orbiting electron equals the frequency 
of the external driving electric microwave field, one finds stable, though 
nonstationary quantum states which are described by nonspreading wave packets 
centered around a classical periodic orbit~\cite{BialynickiBirulaEtAl94,
KalinskiEberly96}. Such Trojan states have first been realized with Lithium 
Rydberg atoms in a linearly polarized microwave field~\cite{MaedaGallagher04}, 
and still are subject of on-going research in atomic physics~\cite{WykerEtAl12}.
More generally, ``Trojan'' single-particle wave packets belong to Floquet 
states which are semiclassically attached to a non-linear resonance island of 
the corresponding classical phase space, thus explaining their non-dispersive 
nature~\cite{HenkelHolthaus92,BuchleitnerEtAl02}.

It has been pointed out recently that Trojan states can also occur in 
periodically driven many-body systems, where they correspond to stable 
collective excitations, or quasiparticles, moving in phase with the driving
force~\cite{GertjerenkenHolthaus14}. Here we show that there exists a genuinely
quantum-mechanical beating effect between similar many-body Trojan states which
perform {\em subharmonic\/} motion with respect to the drive; this beating 
can be understood as quasiparticle tunneling. In particular, we consider a 
resonantly driven bosonic Josephson junction, as provided by ultracold atoms 
in an optically generated double-well potential~\cite{AlbiezEtAl05,GatiMKO07}, 
or by the internal Josephson effect in a spinor Bose-Einstein 
condensate~\cite{ZiboldEtAl10}, and demonstrate that there is a quantum beating
phenomenon between classically equivalent, Trojan-like collective subharmonic 
many-particle motions. Against the background of a mean-field approach this 
beating may be interpreted as dynamical tunneling in the sense of Davis and 
Heller, {\em i.e.\/}, as quantum mechanical tunneling between symmetry-related 
regular regions of classical phase space in the absence of a potential 
barrier~\cite{DavisHeller81,DTBook11}. However, the very core of this beating 
effect reveals itself on the full $N$-particle quantum level, neither taking 
recourse to the mean-field picture nor to the entailing classical phase space, 
and suggests an interesting experimental option for creating and probing 
mesoscopic Schr\"odinger cat states. To lay out our reasoning we first 
sketch the basic mechanism in a quite general, but approximate form in 
Sec.~\ref{sec:2}, shifting technical details of the analysis to the Appendix. 
We then verify our deductions with the help of exact numerical model 
calculations in Sec.~\ref{sec:3}, whereupon some remarks concerning possible 
experimental observations are made in the final Sec.~\ref{sec:4}.

\section{The basic tunneling scheme}
\label{sec:2}

\begin{figure}[t]
\begin{center}
\includegraphics[width = 0.6\linewidth]{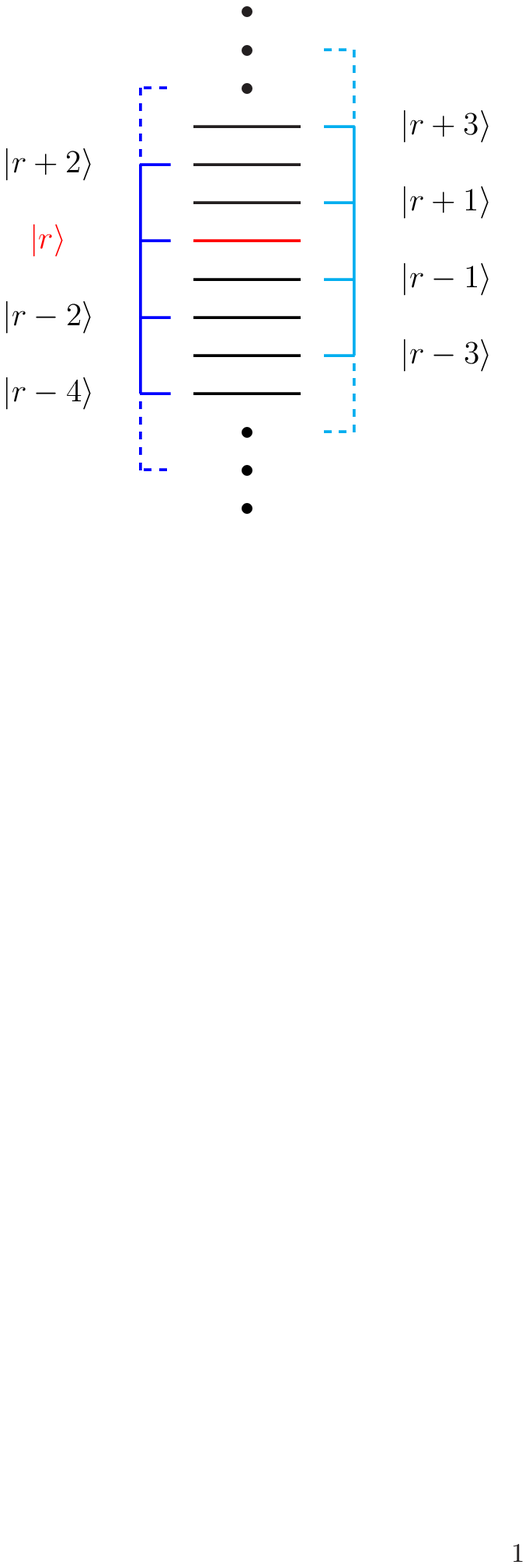}
\end{center}
\caption{(Color online) Principle of Trojan beating for $\nu = 2$: The 
	unperturbed energy eigenstates $|n\rangle$ of a weakly anharmonic 
	oscillator, with energy levels spaced by about $\hbar\omega/2$ in 
	the vicinity of $n = r$ as indicated by the horizontal lines,  are 
	subjected to an external monochromatic perturbation with frequency 
	$\omega$. Resonant coupling then occurs between next-to-nearest
	neighbors, as expressed by Eq.~(\ref{Eq_FRM}), giving rise to two 
	almost uncoupled ``ladders'' of states indicated by the vertical 
	braces. Each ladder corresponds to a set of $2\pi/\omega$-periodic 
	Floquet states enumerated by a new quantum number~$k$. Odd and even
	superpositions of the Floquet states with $k = 0$ yield two 
	$4\pi/\omega$-periodic Trojans, which beat among themselves on much 
	longer timescales.}      
\label{F_1}
\end{figure}

We start by considering a quantum mechanical nonlinear oscillator which may
be given by either a single-particle or a many-particle system, formally 
described by a Hamiltonian $H_0$ with discrete energy eigenvalues $E_n$ and
eigenstates $|n\rangle$, so that $H_0 | n \rangle = E_n | n \rangle$. We 
assume that the eigenvalues are ordered with respect to magnitude and vary
smoothly with $n$ around some particular state $r$, so that it is meaningful
to take the formal derivative 
\begin{equation}
	\widetilde{\omega}_r = \frac{1}{\hbar}
	\left. \frac{\rd E_n}{\rd n} \right|_{n=r} \; ;
\end{equation}
the frequency $\widetilde{\omega}_r$ defined in this manner is the oscillation 
frequency of a wave packet mainly consisting of states in the vicinity of 
$n = r$. We further assume that this oscillator is exposed to an external
influence described by some sinusoidally modulated operator~$V$, such that the 
full Hamiltonian takes the form
\begin{equation}
	H(t) = H_0 + \lambda V \cos\left( \omega t \right) \; ;
\label{Eq_HAM}
\end{equation}
here $\lambda$ is a dimensionless coupling strength. The key point now is that 
the driving frequency be chosen such that $\widetilde{\omega}_r = \omega/\nu$,
meaning that one oscillation cycle of the unperturbed system governed by 
$H_0$ is as long as $\nu$ cycles of the external drive, so that we have a 
$\nu : 1$ resonance. The case $\nu = 1$ applies to the usual 
Trojans~\cite{KalinskiEberly96,BuchleitnerEtAl02,GertjerenkenHolthaus14}; here 
we demand instead that $\nu \ge 2$ be a small integer larger than unity. It is 
then natural to search for solutions to the time-dependent Schr\"odinger 
equation of the form    
\begin{equation}
	| \Psi(t) \rangle = \sum_n c_n(t) |n\rangle 
	\exp\left[-\frac{\ri}{\hbar}
	\left(E_r + \left(n-r\right)\frac{\hbar\omega}{\nu} \right) t 
	\right] \; ,
\label{Eq_ANS}
\end{equation}
where the sum again extends over states close to the resonant state $n = r$.
Because of the resonance condition, the exponential exhibits the first-order
expansions of the energies $E_n$ around $n = r$. This implies that the 
remaining time-dependence of the coefficients $c_n(t)$, given by the exact
system 
\begin{eqnarray}
	\ri\hbar\dot{c}_n(t) & = & 
	\left( E_n - E_r - (n-r)\frac{\hbar\omega}{\nu}\right) c_n(t) 
\\ 	& + &  
	\lambda \cos(\omega t) \sum_m \re^{\ri (n-m)\omega t/\nu}
	\langle n | V |m \rangle c_m(t) \; ,
\nonumber	
\end{eqnarray}
should be relatively weak. Next, we employ three standard 
approximations~\cite{BermanZaslavsky77,Holthaus95}: We expand the energy 
eigen\-values up to second order according to
\begin{equation}
	E_n \approx E_r + (n-r)\frac{\hbar\omega}{\nu}
	+ \frac{1}{2}(n-r)^2 E''_r \; ,
\end{equation}
replace all matrix elements $\langle n | V | n \pm \nu \rangle$ by a 
representative constant~$v$, and keep, in the spirit of the rotating-wave 
approximation, only the secular terms $m = n \pm \nu$, thus being led to the 
strongly simplified system 
\begin{equation}
	\ri\hbar\dot{c}_n = \frac{1}{2}(n-r)^2 E''_r c_n 
	+ \frac{\lambda v}{2} \left( c_{n+\nu} + c_{n-\nu} \right) \; .
\label{Eq_FRM}
\end{equation}
This set of equations contains the essential physics of beating Trojans.
As illustrated in Fig.~\ref{F_1} for the case $\nu = 2$, each coefficient
$c_n$ is coupled only to its remote neighbors $c_{n\pm\nu}$, resulting
in $\nu$ separate subsets of coefficients. The mathematical analysis, carried 
through in detail in the Appendix, shows that this Eq.~(\ref{Eq_FRM}) is 
the Fourier representation of a Mathieu equation, which formally equals 
the stationary Schr\"odinger equation of a fictitious quantum particle moving 
on a one-dimensional cosine lattice, with periodic Born-von Karman boundary 
conditions imposed after $\nu$~potential wells~\cite{AshcroftMermin76}.
This yields Bloch bands of energy eigenstates, labeled by the band index 
$k = 0,1,2,\ldots$, each band containing $\nu$ states. Transformed back to
the nonlinear oscillator considered here, these Bloch states provide solutions
of the form
\begin{eqnarray}
	| \Psi_k^{(j)}(t) \rangle & = & \re^{-\ri\eta_k^{(j)} t/\hbar} 
	\sum_\ell f_{\ell,k}^{(j)} | r + j + \ell\nu \rangle 
\nonumber \\	& \times &	
	\exp\left[ -\frac{\ri}{\hbar}
	\left(E_r + j\hbar\omega/\nu + \ell\hbar\omega\right)t \right] \; ,	
\label{Eq_AFS}
\end{eqnarray}
with $j = 0,1,\ldots,\nu-1$ enumerating the states in the $k$th Bloch band,
$f_{\ell,k}^{(j)}$ denoting the $\ell$th Fourier coefficient of a Mathieu
function specified in the Appendix, and $\eta_k^{(j)}$ being proportional to 
the energy of the Bloch state labeled by $k$ and $j$, thus falling within the 
$k$th energy band.

Importantly, these approximate solutions conform to the Floquet theorem: 
Because the Hamiltonian~(\ref{Eq_HAM}) is periodic in time,  $H(t) = H(t+T)$ 
with $T = 2\pi/\omega$, it gives rise to a complete set of Floquet states, that
is, of {\em exact\/} solutions to the time-dependent Schr\"odinger equation 
which possess the particular form~\cite{Shirley65,Zeldovich67,Sambe73}
\begin{equation}
	| \Psi_n(t) \rangle = | u_n(t) \rangle	 
	\exp(-\ri\varepsilon_n t/\hbar)	
\end{equation}  
with $ | u_n(t) \rangle = | u_n(t+T) \rangle $, and thus reproduce themselves
perpetually in time, except for a phase factor determined by their respective
quasienergy~$\varepsilon_n$. Evidently, the above wave functions~(\ref{Eq_AFS})
are Floquet states with quasienergies 
\begin{equation}
	\varepsilon_k^{(j)} = E_r + j\hbar\omega/\nu + \eta_k^{(j)} 
	\quad (\bmod \; \hbar\omega) \; .
\label{Eq_AQE}
\end{equation}
Hence, within the regime of validity of the above approximations a $\nu : 1$ 
resonance leads to a characteristic ordering of the quasienergy spectrum of a 
driven nonlinear quantum oscillator, featuring $\nu$ sets of Floquet states 
with quasienergies displaced against each other by $\hbar\omega/\nu$. Is is 
assumed here that the depth of the effective cosine lattice, which is 
proportional to the driving strength~$\lambda$, be so large that there are 
several ``below-barrier'' bands. The Floquet states associated with the 
ground-state band $k = 0$ then are of particular interest: For $\nu = 1$, for
which there is $j = 0$ only, the Floquet ground state $k = 0$ provides the 
archetypal Trojan, that is, the maximally localized nonspreading wave packet 
closely following the stable $T$-periodic orbit associated with a $1:1$ 
resonance in classical phase space~\cite{BuchleitnerEtAl02,
GertjerenkenHolthaus14,Holthaus95}; the states with $k \ge 1$ constitute its
excitations.    
  
The next link in the chain of reasoning again is suggested by the solid-state
picture of a quasiparticle moving on a one-dimensional cosine lattice: 
From the extended Bloch waves pertaining to the same energy band of such a 
lattice one can construct Wannier functions which no longer are stationary 
energy eigenstates, but which are localized in the individual potential 
wells~\cite{AshcroftMermin76}. Analogously one may construct, for instance,
the linear combination
\begin{equation}
	| \widetilde{\Psi}_k(t) \rangle = 
	\frac{1}{\sqrt{\nu}} \sum_{j=0}^{\nu-1} | \Psi_k^{(j)}(t) \rangle \; .
\end{equation}      
If all the $\eta_k^{(j)}$ were identical for $j = 0,\ldots,\nu-1$ (that is, 
if the $k$th Bloch band had a vanishing width), this state would correspond 
precisely to the $k$th excitation of a usual Trojan, but now with frequency 
$\omega/\nu$. Hence, it would remain perpetually localized around one of the 
$\nu$ equivalent classical $\nu T$-periodic orbits generated by a $\nu:1$ 
resonance according to the Poincar\'{e}-Birkhoff theorem~\cite{JoseSaletan98}, 
performing coherent motion which is subharmonic with respect to the driving 
frequency. However, quantum tunneling between the individual lattice wells 
bestows a finite width upon the bands, leading to slightly different 
$\eta_k^{(j)}$ and thus to a beating effect between $\nu$ Trojan-like wave 
packets, each following one of these $\nu$ orbits.

\section{Model calculations}
\label{sec:3}

\begin{figure}[t]
\begin{center}
\includegraphics[width = 1.0\linewidth]{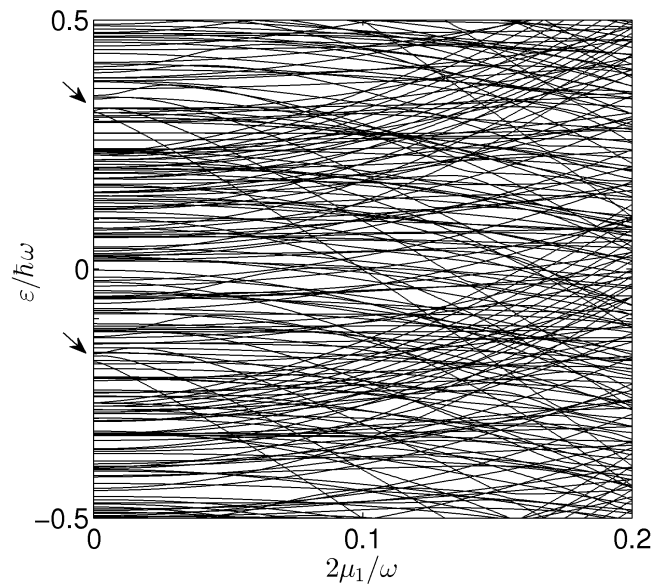}
\end{center}
\caption{Quasienergy spectrum of the tilted, driven bosonic Josephson 
	junction~(\ref{Eq_DBJ}) for $N = 200$ particles, $N\kappa/\Omega=0.95$,
	$\omega/\Omega = 3.258$, and $\mu_0/\Omega = 0.5$. Under these
	conditions  $r = 142$ is a resonant level with $\nu = 2$, leading to 
	two almost identical sets of quasienergies displaced against each 
	other by $\hbar\omega/2$, as predicted by the Mathieu theory. The 
	Trojan doublet with $k = 0$ is indicated by the arrows.} 
\label{F_2}	
\end{figure}

We now apply these considerations to a specific many-body system: The 
unperturbed oscillator is given by the Lipkin-Meshkov-Glick 
Hamiltonian~\cite{LipkinEtAl65}
\begin{eqnarray}
	H_0 & = & -\frac{\hbar\Omega}{2} 
	\left( \ab_1\ad_2 + \ad_1\ab_2 \right) + 
	\hbar\kappa \left( \ad_1\ad_1\ab_1\ab_1 + \ad_2\ad_2\ab_2\ab_2 \right)
\nonumber \\	& & 	
	+ \hbar\mu_0 \left( \ad_1\ab_1 - \ad_2\ab_2 \right)
\label{Eq_LMG}
\end{eqnarray}
which describes a Bose-Einstein condensate in a tilted double-well
potential~\cite{MilburnEtAl97,ParkinsWalls98}. Here the operators
$a^{(\dagger)}_j$ annihilate (create) a Bose particle in well~$j$ 
$(j = 1,2)$, $\hbar\Omega$ is the single-particle tunneling splitting,
$2\hbar\kappa$ denotes the repulsion energy of a pair of bosons occupying the
same well, and $2\hbar\mu_0$ is the energetic misalignment of the two wells.    
We subject this bosonic Josephson junction~(\ref{Eq_LMG}) to an additional 
time-periodic tilt with amplitude $\hbar\mu_1$, such that the total Hamiltonian
reads~\cite{EckardtEtAl05} 
\begin{equation}
	H(t) = H_0 + \hbar\mu_1\sin(\omega t)
	\left( \ad_1\ab_1 - \ad_2\ab_2 \right) \; .
\label{Eq_DBJ}
\end{equation}
In Fig.~\ref{F_2} we plot the exact, numerically computed quasienergies 
of this system for $N = 200$ particles, scaled interaction strength  
$N\kappa/\Omega=0.95$, scaled driving frequency $\omega/\Omega = 3.258$, 
and scaled tilt $\mu_0/\Omega = 0.5$ vs.\ the scaled driving strength 
$2\mu_1/\omega$. Here $r = 142$ is a resonant level with $\nu = 2$, leading 
in accordance with Eq.~(\ref{Eq_AQE}) to two almost identical sets of 
quasienergies displaced against each other by $\hbar\omega/2$. As shown in
the Appendix, the Mathieu theory yields a quite good description of these 
quasienergies, notwithstanding the somewhat crude approximations made, so 
that the analysis sketched in the previous section provides a sound basis    
for our deductions.

\begin{figure}[t]
\begin{center}
\includegraphics[width = 1.0\linewidth]{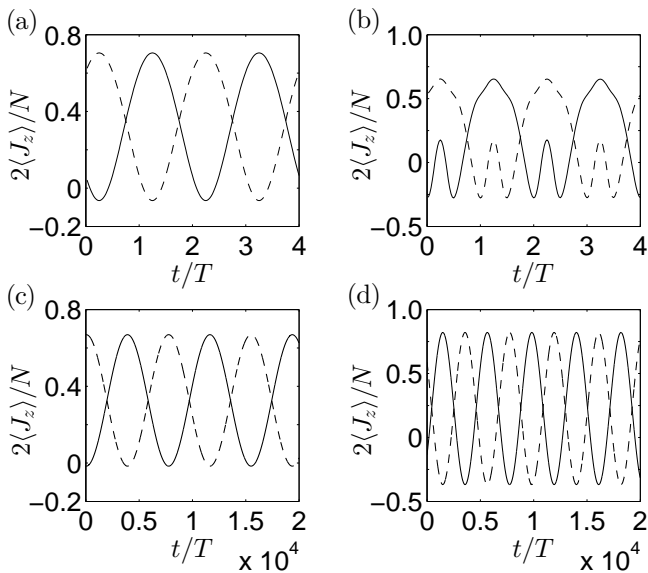}
\end{center}
\caption{Scaled population imbalance $2\langle J_z \rangle/N$ evolving from 
	the Trojan initial states $|\widetilde{\Psi}_0^{(+)}\rangle$ (full 
	lines) and $|\widetilde{\Psi}_0^{(-)}\rangle$ (dashed lines). While 
	the short-time evolution is monitored continuously, the long-time
	evolution is recorded stroboscopically at each multiple of $2T$. 
	(a) and (c): $N = 100$, $2\mu_1/\omega = 0.025$, the other parameters 
	are as in Fig.~\ref{F_2}. (b) and (d): $N = 50$, $2\mu_1/\omega = 1.0$, 
	$N\kappa/\Omega = 0.95$, $\omega/\Omega = 1.6$, $\mu_0/\Omega = 0.8$.} 	 
\label{F_3}
\end{figure}

Next, we take the numerically computed exact superpositions
\begin{equation}
	| \widetilde{\Psi}_0^{(\pm)}(0) \rangle  = 
	\left( | \Psi_0^{(0)}(0) \rangle 
	\pm | \Psi_0^{(1)}(0) \rangle \right) / \sqrt{2}
\label{Eq_TQP}
\end{equation}
and follow their time evolution by solving the time-dependent $N$-particle
Schr\"odinger equation: These states correspond to stable,
Trojan-like collective many-particle excitations which have been termed
``floton'' quasiparticles in Ref.~\cite{GertjerenkenHolthaus14}; they should 
beat among themselves on a time scale determined by the tiny difference 
between the energies $\eta_0^{(0)}$ and $\eta_0^{(1)}$. Figure~\ref{F_3} shows 
results of such calculations, for both short and long times. Here we plot the 
experimentally measurable population difference $2 \langle J_z \rangle /N$ 
between both wells, with $J_z = (\ad_1\ab_1 - \ad_2\ab_2)/2 $. We first 
take $N = 100$ particles and fix the driving amplitude at the small value 
$2\mu_1/\omega = 0.025$, leaving the other parameters as in Fig.~\ref{F_2}. 
Over short intervals we then merely observe the coherent $2T$-periodic 
population exchange, as depicted in Fig.~\ref{F_3} (a), while the expected 
Trojan beating manifests itself if we plot the population difference 
stroboscopically at multiples of $2T$ only, but for much longer durations, 
as done in Fig.~\ref{F_3} (c). Here the observed Trojan tunneling time 
$T_0^{({\rm tun})}$, {\em i.e.\/}, the duration of half a beating cycle, 
is $3865 \, T$. For comparison, the Mathieu estimate~(\ref{SM_EST}) derived 
in the Appendix gives $T_0^{({\rm tun})} \approx 1.9 \times 10^3 \, T$, and 
thus already provides the correct order of magnitude. We stress that the 
beating effect found here should not be attributed to the tunneling of 
individual Bose particles in the physical double-well potential constituting 
the Josephson junction, but rather to the tunneling of a single quasiparticle 
in the effective double cosine well in Fourier space. Quite similar observations
are also made for significantly stronger driving: The tunneling signatures 
recorded in Figs.~\ref{F_3} (b) and (d) emerge for $2\mu_1/\omega = 1.0$.

\begin{figure}[t]
\begin{center}
\includegraphics[width = 1.0\linewidth]{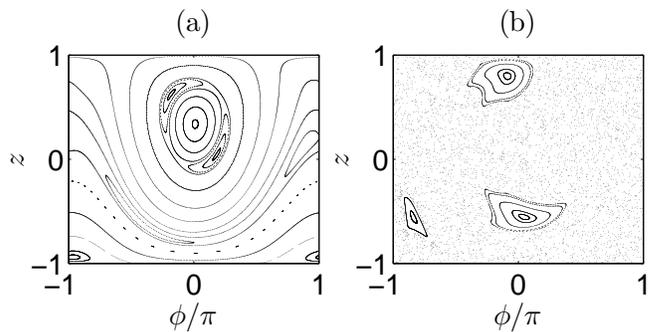}
\end{center}
\caption{Poincar\'{e} surfaces of section generated by the classical driven 
	pendulum~(\ref{Eq_HDP}). (a) For the parameters underlying 
	Fig.~\ref{F_3} (a) and (c), the $2:1$ resonance leads to two
	banana-shaped zones surrounding the central elliptic fixed point.   	  
        (b) For the parameters of Fig.~\ref{F_3} (b) and (d), the two 
	resonant islands are separated by a chaotic sea; an additional
	regular island is visible in the lower left corner.}  
\label{F_4}
\end{figure}

Within a mean-field ansatz, involving the introduction of a macroscopic 
wave function and the uncontrolled factorization of expectation values of 
operator products into products of expectation values~\cite{Leggett01}, the 
dynamics of $N$~Bose particles in the driven Josephson junction~(\ref{Eq_DBJ})
are reduced to those of merely two amplitudes describing the population of 
the two wells. In terms of the population difference~$z$ and the relative 
phase~$\phi$ which adopt the roles of momentum and its conjugate position 
coordinate, respectively, the mean-field dynamics coincide exactly with 
those of the driven, nonrigid classical pendulum governed by the Hamiltonian 
function~\cite{SmerziEtAl97,WeissTeichmann08}    
\begin{eqnarray}
	H_{\rm{mf}}(z,\phi,t) & = & N\kappa z^2 
	- \Omega \sqrt{1-z^2}\cos(\phi) 
\nonumber \\ & & 	
	+ 2z\Big(\mu_0 + \mu_1\sin(\omega t) \Big) \; .
\label{Eq_HDP}
\end{eqnarray}
In Fig.~\ref{F_4} we show Poincar\'e sections for this classical nonlinear 
pendulum~\cite{JoseSaletan98}, obtained by recording typical trajectories
$T$-stroboscopically in the $\phi$-$z$-plane. For the parameters chosen in 
Fig.~\ref{F_3} (a) and (c) for illustrating Trojan tunneling, the $2:1$ 
resonance manifests itself as two banana-shaped zones surrounding the central 
elliptic fixed point within a practically regular phase space, whereas the 
resonant islands are separated by a chaotic sea for the strong-driving scenario
considered in Fig.~\ref{F_3} (b) and (d). In both cases the Trojan states 
$k = 0$ are semiclassically associated with the innermost ``quantized'' closed 
contours $\gamma_k$ encircling the resonant stable $2T$-periodic orbits which 
are selected by the Einstein-Brillouin-Keller 
conditions~\cite{GertjerenkenHolthaus14}
\begin{equation}
	\frac{1}{2\pi} \oint_{\gamma_k} \! z \rd \phi = 
	\frac{2}{N}\left( k + \frac{1}{2} \right) \; ;
\label{Eq_EBK}	
\end{equation}
observe that $2/N = \hbar_{\rm eff}$ here plays the role of an effective 
Planck constant. In the classical case, as corresponding to the mean-field
approximation to the full $N$-particle dynamics, a trajectory starting
in one of the two equivalent resonant islands inevitably ends up in the
other one after one period~$T$, and returns to the first island after $2T$.
The quasiparticle tunneling discussed in this work is a beyond-mean-field 
effect which can be regarded as a form of dynamical tunneling between 
symmetry-related regular regions of phase space~\cite{DavisHeller81,DTBook11}: 
After the Trojan tunneling time $T_0^{({\rm tun})}$ the $N$-particle state is 
semiclassically associated with the ``wrong'' island.

\section{Discussion}
\label{sec:4}

Since single-particle Trojan states with $ \nu = 1$ could be observed over 
about 15\,000 cycles in microwave-driven Rydberg atoms~\cite{MaedaGallagher04},
it seems feasible to detect Trojan beating for $\nu \ge 2$ with highly excited 
atoms in suitably tuned microwave fields. This already would constitute a 
spectacular demonstration of a genuine quantum effect. The search for Trojan 
many-body beating in driven bosonic Josephson junctions might break even 
further ground. Exerimentally, one could generate Trojans in a robust manner
by means of adiabatic switching~\cite{GertjerenkenHolthaus14}, starting from 
the ground state. Because the population imbalance is a well-accessible
observable, Trojan states with $\nu = 2$ can be identified through their
subharmonic motion. The signature of Trojan quasiparticle tunneling then
would be striking: After the tunneling time, the sloshing condensate is
out of phase with the drive, being found in the ``wrong'' well. Since the 
Trojan quasiparticles~(\ref{Eq_TQP}), or those with even higher~$\nu$, 
correspond to highly entangled states, measurements of Trojan beating for 
different particle numbers could enable one to gain valuable  information on 
the persistence (or non-persistence) of entanglement with increasing~$N$ 
after long driving times. 

This many-body aspect distinguishes our proposal from related recent works 
which have addressed dynamical tunneling with Bose-Einstein condensates in 
magnetic microtraps~\cite{LenzEtAl13}, or in a kicked-rotor 
experiment~\cite{ShresthaEtAl13}: In such configurations one probes essentially
single-particle physics, so that the effective Planck constant is determined 
by the parameters of the respective set-up~\cite{LenzEtAl13,ShresthaEtAl13}. 
In contrast, Eq.~(\ref{Eq_EBK}) has shown that $\hbar_{\rm eff} = 2/N$ in our
case, so that here the ``degree of quantumness'' can be tuned by varying 
the particle number. This might be of interest for exploring quasiparticle 
tunneling in quasi-regular situations exemplified by Fig.~\ref{F_4} (a), or 
in chaotic situations as corresponding to Fig.~\ref{F_4} (b), and for testing
modern concepts of chaos-assisted tunneling~\cite{LockEtAl10} within a 
many-body setting.    
    
The experimental exploitation of Trojan quasiparticles presupposes that the 
periodically driven Bose-Einstein condensate system possesses a well-preserved 
order parameter, in order to render the existence of Floquet condensates 
possible. This requirement puts an upper limit on the sizes of the systems 
one could work with, since Floquet condensates tend to become unstable upon 
increasing~$N$~\cite{GertjerenkenHolthaus14,GertjerenkenHolthaus14b}.
It is an open question whether one could detect signs of the onset of this
instability in Trojan quasiparticle tunneling.

\begin{acknowledgments}
We acknowledge support from the Deutsche Forschungsgemeinschaft (DFG) through 
grant No.\ HO 1771/6-2. The computations were performed on the HPC cluster 
HERO, located at the University of Oldenburg and funded by the DFG through
its Major Research Instrumentation Programme (INST 184/108-1 FUGG), and by 
the Ministry of Science and Culture (MWK) of the Lower Saxony State.
\end{acknowledgments}

\appendix

\section{Mathieu analysis of beating Trojans}

The starting point is the approximate system~(\ref{Eq_FRM}): Because each 
coefficient $c_n$ is coupled to $c_{n\pm\nu}$ only, there are $\nu$ separate 
``ladders'' of states (see Fig.~\ref{F_1} for $\nu = 2$) which we label by 
$j = 0,1,\ldots,\nu-1$. Accordingly, we relabel the coefficients such that
\begin{equation}
	c_{r + j + \ell\nu} \equiv b_{\ell}^{(j)} \; ,
\end{equation}
so that the index $\ell$ enumerates the members of the particular subset of
coefficients specified by~$j$. Since this implies
\begin{equation}
	(n - r)^2 = \nu^2(\ell + j/\nu)^2 \; ,	
\end{equation}
we have $\nu$ uncoupled systems
\begin{equation}
	\ri\hbar \dot{b}_\ell^{(j)} = \frac{1}{2}\nu^2 E''_r 
	\left( \ell + \frac{j}{\nu} \right)^2 b_\ell^{(j)} 
	+ \frac{\lambda v}{2} 
	\left( b_{\ell+1}^{(j)} + b_{\ell-1}^{(j)} \right) \; . 
\label{SM_FRM}
\end{equation}
Let us first reconsider the standard case $\nu = 1$~\cite{BermanZaslavsky77,
Holthaus95}, meaning that there is only one ladder, $j = 0$. Then the above
Eq.~(\ref{SM_FRM}) is a Fourier representation of the well-known Mathieu
equation~\cite{AbramowitzStegun72}, which plays a central role in both the 
stability analysis of the Paul trap~\cite{Paul90}, and, closely related,
in the analysis of the original Trojan wave packets~\cite{KalinskiEberly96}:
Setting
\begin{equation}
	b_\ell^{(0)}(t) = \re^{-\ri\eta t/\hbar} \frac{1}{2\pi} 
	\int_{0}^{2\pi} \! \rd \vartheta \, f(\vartheta) 
	\re^{-\ri\ell\vartheta} \; ,
\label{SM_COE}
\end{equation}
we immediately have
\begin{equation}
	\ri\hbar \dot{b}_\ell^{(0)} = \eta b_\ell^{(0)}	
\label{SM_AE1}
\end{equation}
and
\begin{equation}
	b_{\ell+1}^{(0)} + b_{\ell-1}^{(0)} =
	\re^{-\ri\eta t/\hbar} \frac{1}{2\pi}
	\int_{0}^{2\pi} \! \rd \vartheta \, f(\vartheta)
	\re^{-\ri\ell\vartheta} 2 \cos\vartheta \; .	
\label{SM_AE2}
\end{equation}
Moreover,
\begin{eqnarray}
	\ell^2 b_\ell^{(0)} & = & \re^{-\ri\eta t/\hbar} \frac{1}{2\pi}
	\int_{0}^{2\pi} \! \rd \vartheta \, f(\vartheta)
	\left( -\frac{\rd^2}{\rd\vartheta^2} \right) \re^{-\ri\ell\vartheta}
\nonumber \\	& = &
	\re^{-\ri\eta t/\hbar} \frac{1}{2\pi}
	\int_{0}^{2\pi} \! \rd \vartheta \, \big( -f''(\vartheta) \big)
	\re^{-\ri\ell\vartheta}	\; ,
\label{SM_PIN}
\end{eqnarray}
provided that $f(\vartheta) = f(\vartheta + 2\pi)$, so that the partial
integrations carried out here do not produce boundary terms. Thus, for 
$\nu = 1$ Eq.~(\ref{SM_FRM}) transforms into 
\begin{equation} 
	\eta f(\vartheta) = -\frac{1}{2}E''_r f''(\vartheta)
	+ \lambda v \cos\vartheta f(\vartheta)
\end{equation}
which is a Mathieu equation; substituting $\vartheta = 2z$ and writing
$f(2z) \equiv \chi(z)$ produces its standard form~\cite{AbramowitzStegun72}
\begin{equation}
	\left[ \frac{\rd^2}{\rd z^2} + \alpha - 2q \cos(2z) \right] \chi(z)
	= 0
\label{SM_SFM}
\end{equation}	
with parameters
\begin{eqnarray}
	\alpha & = & \frac{8\eta}{E''_r} \; ,
\label{SM_MPA}	\\
	q & = & \frac{4\lambda v}{E''_r} \; .	
\end{eqnarray}
Because of the periodic boundary condition $f(\vartheta) = f(\vartheta + 2\pi)$
imposed on $f$ in order to guarantee the validity of Eq.~(\ref{SM_PIN})
we require $\pi$-periodic Mathieu functions $\chi(z) = \chi(z+\pi)$, which
exist only if the parameter $\alpha$ adopts, for given~$q$, one of the 
so-called characteristic values. Employing the widely accepted symbols $a_n$ 
(giving even Mathieu functions) and $b_n$ (giving odd ones) as defined in 
Ref.~\cite{AbramowitzStegun72} for these quantities, the allowed values 
of $\alpha$ are
\begin{equation}
	\alpha_k = \left\{ \begin{array} {ll}
		a_k 	& , \; k = 0,2,4,\ldots \\	
		b_{k+1} & , \; k = 1,3,5,\ldots \; ,
	\end{array} \right.
\label{SM_ALK}
\end{equation}   
thus introducing a new quantum number~$k$. This specifies the desired 
coefficients~(\ref{SM_COE}) as 
\begin{equation}
	b_\ell^{(0)}(t) = \re^{-\ri\eta_k t/\hbar} f_{\ell,k} \; ,
\end{equation}
writing
\begin{equation}
	\eta_k = \frac{1}{8} \alpha_k E''_r
\end{equation}	 
in accordance with Eq.~(\ref{SM_MPA}), and denoting by $f_{\ell,k}$ the 
$\ell$th Fourier coefficient of the Mathieu function associated with 
$\alpha_k$. In this way we have found approximate solutions to the 
time-dependent Schr\"odinger equation of the driven nonlinear oscillator 
which conform to the ansatz~(\ref{Eq_ANS}) made in Sec.~\ref{sec:2}: 
\begin{equation}
	| \Psi_k(t) \rangle = \re^{-\ri\eta_k t/\hbar} \sum_\ell f_{\ell,k}
	| r + \ell \rangle \exp\left[
	-\frac{\ri}{\hbar}\left(E_r + \ell\hbar\omega\right)t \right] \; .	
\label{SM_TWP}
\end{equation}
Obviously these solutions are Floquet states with quasienergies 
\begin{equation}
	\varepsilon_k = E_r + \eta_k \quad (\bmod \; \hbar\omega) \; ;
\end{equation}
the ``ground state'' with $k = 0$ then corresponds to the nonspreading Trojan 
wave packet most strongly localized around the classical periodic orbit which
is locked to the periodic drive in a $1:1$ resonance~\cite{Holthaus95}.

The task now is to generalize this procedure, which so far applies to $\nu = 1$ 
only, to the case $\nu \ge 2$ which is a precondition for Trojan tunneling. 
If we started again from a representation~(\ref{SM_COE}) for $b_\ell^{(j)}(t)$, 
with $f(\vartheta)$ appropriately replaced by $f^{(j)}(\vartheta)$, the analogs
of Eqs.~(\ref{SM_AE1}) and (\ref{SM_AE2}) would go through unchanged, but 
Eq.~(\ref{SM_PIN}) is of no use when $j \neq 0$. Instead, with a view towards 
Eq.~(\ref{SM_FRM}) we write
\begin{equation}
	b_\ell^{(j)}(t) = \re^{-\ri\eta t/\hbar} \frac{1}{\nu 2\pi} 
	\int_{0}^{\nu 2\pi} \! \rd \vartheta \, g^{(j)}(\vartheta) \,  
	\re^{-\ri(\ell + j/\nu)\vartheta} \; ,	
\end{equation}
where 
\begin{equation}
	g^{(j)}(\vartheta) = f^{(j)}(\vartheta) \, 
	\re^{\ri(j/\nu)\vartheta}
\label{SM_RBW}		
\end{equation}
now obeys the crucial boundary condition
\begin{equation}
	g^{(j)}(\vartheta) = g^{(j)}(\vartheta + \nu2\pi)
\label{SM_RBC}
\end{equation}
which is required for establishing the identity
\begin{eqnarray} & &   
	\left(\ell + \frac{j}{\nu}\right)^2 b_\ell^{(j)} 
\nonumber \\ & = & 	 	
	\re^{-\ri\eta t/\hbar}   		
	\frac{1}{\nu2\pi}
	\int_{0}^{\nu2\pi} \! \rd \vartheta \, 
	\big( -g^{(j)}{''}(\vartheta) \big)
	\re^{-\ri(\ell+j/\nu)\vartheta}	\; . \phantom{XXX}
\end{eqnarray}	
This then leads to the Mathieu equation
\begin{equation}
	\eta g^{(j)}(\vartheta) = -\frac{\nu^2}{2}E''_r g^{(j)}{''}(\vartheta)
	+ \lambda v \cos\vartheta g^{(j)}(\vartheta)	
\label{SM_RME}
\end{equation}
which again can be brought into the standard form~(\ref{SM_SFM}) by setting
$g^{(j)}(2z) \equiv \chi(z)$, implying
\begin{eqnarray}
	\alpha & = & \frac{8\eta}{\nu^2 E''_r} \; ,
\\
	q & = & \frac{4\lambda v}{\nu^2 E''_r} \; .	
\end{eqnarray}
The key point now is the boundary condition~(\ref{SM_RBC}), which forces us 
to select $\nu\pi$-periodic Mathieu functions $\chi(z) = \chi(z+\nu\pi)$.
Because Eq.~(\ref{SM_RME}) can be interpreted as the stationary Schr\"odinger
equation for a quantum particle moving on a one-dimensional cosine lattice, 
the associated desired solutions~(\ref{SM_RBW}) then correspond to Bloch waves 
of this lattice, with periodic Born-von Karman boundary conditions imposed 
after $\nu$ potential wells~\cite{AshcroftMermin76}. We denote the discrete 
Bloch band index labeling these solutions by~$k = 0,1,2,\ldots\,$, in 
accordance with the notation employed in Eq.~(\ref{SM_ALK}) for $\nu = 1$. 
Thus, for each~$k$ we find the $\nu$ different approximate solutions to the 
time-dependent Schr\"odinger equation of the driven nonlinear oscillator
which have been heralded by Eq.~(\ref{Eq_AFS}), each one corresponding to a 
``ladder'' of the type depicted in Fig.~\ref{F_1}: 
\begin{eqnarray}
	| \Psi_k^{(j)}(t) \rangle & = & \re^{-\ri\eta_k^{(j)} t/\hbar} 
	\sum_\ell f_{\ell,k}^{(j)} | r + j + \ell\nu \rangle 
\nonumber \\	& \times &	
	\exp\left[ -\frac{\ri}{\hbar}
	\left(E_r + j\hbar\omega/\nu + \ell\hbar\omega\right)t \right] \; .	
\end{eqnarray}
Once again these solutions are Floquet states, now with quasienergies 
\begin{equation}
	\varepsilon_k^{(j)} = E_r + j\hbar\omega/\nu + \eta_k^{(j)} 
	\quad (\bmod \; \hbar\omega) \; .
\end{equation}
In classical mechanics a $\nu : 1$ resonance gives rise to $\nu$ equivalent
stable $\nu \times 2\pi/\omega$-periodic orbits~\cite{JoseSaletan98}. The
above $2\pi/\omega$-periodic Floquet states, corresponding to extended Bloch 
waves in a lattice with $\nu$ wells, yield wave packets localized around 
{\em each\/} of these orbits, the sharpest localization being obtained for 
$k = 0$. In order to construct wave packets which follow {\em only one\/} 
orbit in a Trojan fashion, one has to take those linear combinations of the 
Bloch waves which produce a Wannier-like state localized in an individual 
well~\cite{AshcroftMermin76}. In contrast to single Floquet states such 
linear combinations are no stationary states, which is the reason for Trojan
beating.

\begin{figure}
\begin{center}
\includegraphics[width = 1.0\linewidth]{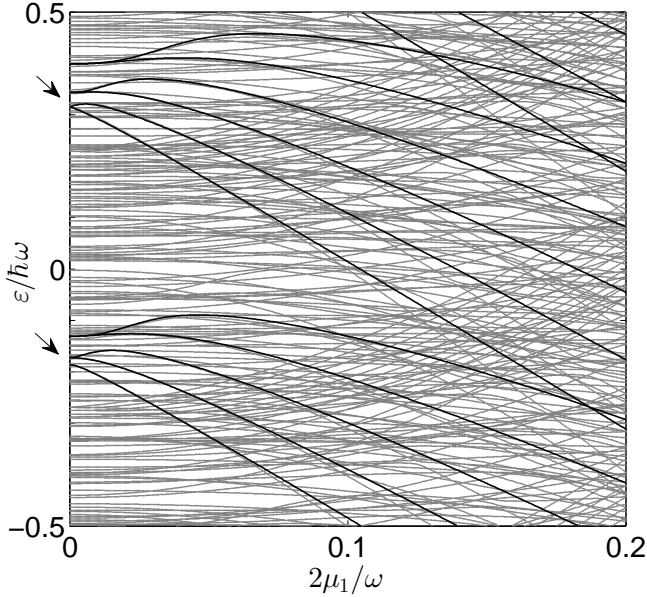}
\end{center}
\caption{(Color online) Grey lines: Exact quasienergy spectrum of the tilted, 
	driven bosonic Josephson junction for $N = 200$ particles, as also 
	shown in Fig.~\ref{F_2} ($N\kappa/\Omega=0.95$, $\omega/\Omega = 3.258$,
	$\mu_0/\Omega = 0.5$). Black lines: Quasienergies provided for 
	$\nu = 2$ by the Mathieu approximations~(\ref{SM_MA1}) and
	(\ref{SM_MA2}), with $r = 142$. The Mathieu parameter~(\ref{SM_MAP})
	is given by $q\omega/(2\mu_1) = -498.26$.}    
\label{F_Supl}	
\end{figure}

All essentials are made visible already by the case $\nu = 2$, which leads 
to a lattice with periodic boundary conditions imposed after two wells: Then
one encounters the tunneling effect of a quasiparticle in a symmetric 
double-well potential, involving the superpositions
\begin{equation}
	| \widetilde{\Psi}_k^{(\pm)} \rangle = \frac{1}{\sqrt{2}}
	\left( 
	| \Psi_k^{(0)} (t=0) \rangle \pm| \Psi_k^{(1)} (t=0) \rangle 
	\right)  
\end{equation}      
which correspond to states initially localized in the ``left'' or ``right'' 
well, respectively. Here the ``ground state doublet'' $k=0$ yields two 
strongly localized wave packets, each following closely, on short time scales, 
one of the two stable periodic orbits associated with the $2:1$  resonance. 
While the quasienergies for $j = 0$ take the form  
\begin{equation} 
	\varepsilon_k^{(0)} = E_r + \frac{1}{2}\alpha_k^{(0)}E''_r 
	\quad (\bmod \; \hbar\omega) \; , 	
\label{SM_MA1}
\end{equation}
with $\alpha_k^{(0)}$ being given by Eq.~(\ref{SM_ALK}), one obtains  
\begin{equation} 
	\varepsilon_k^{(1)} = E_r + \frac{1}{2}\hbar\omega + 
	\frac{1}{2}\alpha_k^{(1)}E''_r \quad (\bmod \; \hbar\omega)  
\label{SM_MA2}
\end{equation}
for $j = 1$, now requiring the Mathieu characteristic values associated with
$2\pi$-periodic Mathieu functions $\chi(z)$~\cite{AbramowitzStegun72}:
\begin{equation}
	\alpha_k^{(1)} = \left\{ \begin{array} {ll}
		b_{k+1} & , \; k = 0,2,4,\ldots \\	
		a_k 	& , \; k = 1,3,5,\ldots \; .
	\end{array} \right.
\end{equation}   
Evidently, the Trojan tunneling time --- that is, the time to evolve from 
$| \widetilde{\Psi}_k^{(+)} \rangle$ to $| \widetilde{\Psi}_k^{(-)} \rangle$, 
or to tunnel from one of the two $4\pi/\omega$-periodic orbits to the other 
--- is given by
\begin{equation}
	T_k^{({\rm tun})} = \frac{2\pi\hbar}{E''_r(b_{k+1} - a_k)} \; ,
\end{equation}
which can be evaluated for large~$q$ by employing the asymptotic 
expression~\cite{AbramowitzStegun72,MeixnerSchaefke54}
\begin{equation}
	b_{k+1} - a_k \sim \frac{2^{4k+5}}{k!} \sqrt{\frac{2}{\pi}} \, 
	q^{k/2 + 3/4} \re^{-4\sqrt{q}} \; . 
\end{equation}
In particular, for the archetypal Trojan doublet with $k = 0$ one obtains
the estimate
\begin{equation} 
	T_0^{({\rm tun})} \sim \frac{\pi^{3/2}\hbar}{16\sqrt{2} E''_r}
	q^{-3/4} \re^{4\sqrt{q}} \; . 
\label{SM_EST}
\end{equation}
As is evident from the standard form~(\ref{SM_SFM}), the Mathieu parameter~$q$
determines the depth of the cosine lattice. In the case of the resonantly 
driven bosonic Josephson junction defined by Eqs.~(\ref{Eq_LMG}) and 
(\ref{Eq_DBJ}) in Sec.~\ref{sec:3}, it is calculated from 
\begin{equation}
	q = \frac{2}{\nu^2 E''_r/(\hbar\omega)} \frac{2\mu_1}{\omega}
	\langle r | a^\dagger_1 a^{\phantom{\dagger}}_1 -
	a^\dagger_2 a^{\phantom{\dagger}}_2 | r - \nu \rangle \; .
\label{SM_MAP}
\end{equation}
Figure~\ref{F_Supl} provides a comparison of the exact quasienergies for the 
$2:1$~resonance already considered in Fig.~\ref{F_2} to the corresponding 
approximations~(\ref{SM_MA1}) and (\ref{SM_MA2}) provided by the Mathieu 
analysis. In view of the substantial simplifications which have led to its 
starting point~(\ref{SM_FRM}), the agreement is quite satisfactory, confirming
that the essential features have been kept.

\end{document}